\documentclass{jpsj-suppl}
\usepackage{txfonts} 

\title{Penta-quark States with Strangeness, Hidden Charm and Beauty}

\author{Jia-Jun \textsc{ Wu}$^{1}$ and Bing-Song \textsc{Zou}$^{2}$}

\inst{$^{1}$Special Research Center for the Subatomic Structure of Matter (CSSM),
School of Chemistry and Physics, University of Adelaide, Adelaide 5005, Australia \\
$^{2}$State Key Laboratory of Theoretical Physics, Kavli Institute
for Theoretical Physics China, Institute of Theoretical Physics,
Chinese Academy of Sciences, Beijing 100190,China}

\email{jiajun.wu@adelaide.edu.au}

\recdate{August 30, 2015}

\abst{
The classical quenched quark models with three constituent quarks
provide a good description for the baryon spatial ground states, but
fail to reproduce the spectrum of baryon excited states.
More and more evidences suggest that unquenched effects with
multi-quark dynamics are necessary ingredients to solve the problem.
Several new hyperon resonances reported recently could fit in the picture
of penta-quark states. Based on this picture, some new hyperon
excited states were predicted to exist; meanwhile with extension
from strangeness to charm and beauty, super-heavy narrow $N^*$ and
$\Lambda^*$ resonances with hidden charm or beauty were predicted to
be around 4.3 and 11 GeV, respectively. Recently, two of such $N^*$
with hidden charm might have been observed by the LHCb experiment.
More of those states are expected to be observed in near future.
This opens a new window in order to study hadronic dynamics for the
multi-quark states.}

\kword{Penta-quark, Hadron spectroscopy, Multi-quark state}

\begin{document}
\maketitle

\section{Introduction}

Correct description of hadron spectrum is one of the most important
ways to understand the strong interaction governed by the Quantum
chromodynamics (QCD) theory.
For example, through the discovery of ground baryons, such as
proton, neutron, $\Delta(1232)$, $\Sigma$, $\Xi$ and so on, the
classical three-quark constituent model was built to describe
various ground baryons, and successfully predicted $\Omega(sss)$
baryon which was confirmed by later experiments with the mass around
1670 MeV.
In classical constituent quark models, hadrons are ascribed as
baryon composed of three-quarks and meson composed of a
quark-antiquark pair.
In these models, ground states are that each quark is in S-wave of
orbital angular momentum ($L=0$) and radial ground state ($n=0$),
while excited states are those with at least one of the quantum numbers of $(L, n)$
is larger than 0.
Thus the lowest excitation of baryon is $(L, n) = (1, 0)$, and their
total spin and parity are $J^p = \frac{1}{2}^-$.

%
As we known, even for the lowest excited hadron states, spectra
predicted from the classical quark models are not consistent with
observations of various experiments.
For example, the lowest negative parity baryons are $N^*(1535)$ and
$\Lambda^*(1405)$ with $J^p=\frac{1}{2}^-$~\cite{pdg}.
However, the three-quark model expects that $N^*(1535)$ with
$|uud\rangle$ is lighter than $\Lambda^*$ with $|uds\rangle$.
Furthermore, it also expects $N^*(1535)$ is lighter than $N^*(1440)$
which is considered as the first radially state of nucleon with $J^p =
\frac{1}{2}^+$.
In fact, the $N^*(1535)$ is heavier than both $\Lambda^*(1405)$ and
$N^*(1440)$.
This is the long-standing mass order reverse problem for the three lowest excited baryons.
For the meson sector, the same thing is happening.
The scalar nonet with $J^p = 0^+$ including $f_0(500)$,
$\kappa(600-800)$, $a_0(980)$ and $f_0(980)$, is the lowest excited
meson nonet ($L=1$).
However, $f_0(500)$, $\kappa(600-800)$ and $f_0(980)$ are all
lighter than their vector partner, $\omega(782)$, $K^*(892)$ and
$\phi(1020)$.
In the classical $q\bar{q}$ model, $f_0(500)$ and $a^0_0(980)$ are
regarded as $(u\bar{u}+d\bar{d})/\sqrt{2}$ and
$(u\bar{u}-d\bar{d})/\sqrt{2}$,  while $f_0(980)$ is $s\bar{s}$
state.
As a result, this model cannot explain why the mass of $a^0_0(980)$
degenerates with $f_0(980)$ rather than close to $f_0(500)$.
In summary, the classical constituent quark models cannot explain
the excited hadron spectrum.
It implies new components are needed to describe hadrons.
In this paper,  we focus on the baryon sector.

In the QCD field,  the multi-quark component is not forbidden
because $q\bar{q}$ can be dragged out from the glue field, in other
words, the number of constituent quarks in a hadron is not a
constant.
This new picture can be named as unquenched quark model, while
original picture is quenched quark model.
Since multi-quark states are available in unquenched quark models,
the baryon spectrum predicted from these models will be very
different from that of three-quark models.
Furthermore, these multi-quark components can naturally explain  the
problem in the three-quark model.
For instance, the mass order reverse problem is easily solved by
including large penta-quark components in these baryon states
\cite{Liu:2005pm, 5qRiska, zhang, Santopinto:2014opa }.

The $N^*(1535)$, $\Lambda^*(1405)$ and $N^*(1440)$ could have large
$|udus\bar{s}>$, $|udsq\bar{q}>$ ($q$=$u$ or $d$) and
$|uduq\bar{q}>$ components, respectively.
The higher mass of $N^*(1535)$ is due to its large $s\bar{s}$
component.
On the other hand, the $s\bar{s}$ component also results in its
large couplings to the channels with strangeness, such as $N\eta$,
$N\eta'$, $N\phi$ and $K\Lambda$.
By recent experimental data and theoretical analyses, $N^*(1535)$
is indeed strongly coupled with these channels~\cite{pdg,
Liu:2005pm, Geng:2008cv, Dugger:2005my, Cao:2008st, Xie:2007qt,
Cao:2009ea}.

There are two possible ways to form penta-quark states: colored and
uncolored quark cluster.
The former one considers the five-quark as
diquark-diquark-antiquark~\cite{zou5q, Jaffe:1976ig, zounul}, where
diquark is a $qq$ colored cluster;
The latter one corresponds to the meson-baryon coupled channel model
~\cite{Kaiser:1995cy, Oset:1997it, Oller:2000fj, Inoue:2001ip,
GarciaRecio:2003ks,  Hyodo:2002pk, Tornqvist:1984fy, Ono:1983rd,
Kalashnikova:2005ui, Pennington:2007xr, Ferretti:2012zz,
Ferretti:2013faa, Ferretti:2013vua, Bijker:2009up, Bijker:2012zza},
for example, $\Lambda^*(1405)$ can be dynamically generated from the
coupled channels of $\bar{K}N$ and $\Sigma\pi$~\cite{oller}.

Both penta-quark and three-quark components may exist in baryons.
For $J^p=\frac{1}{2}^-$ baryons, to excite a constituent quark to be
L=1 state and to drag out a light $q\bar{q}$ pair from gluon field,
the two different mechanisms may be comparable.
%
Therefore, these baryons are possibly the mixtures of the
three-quark  and five-quark components.
Then a very natural question is that how the three-quark and
five-quark configurations transfer to each other ?
The breathing model is proposed in Refs.\cite{zounul,t5q3qa}, here,
the mass of the lowest $\Omega^*$ is predicted by including  the
$sss \leftrightarrow sssq\bar{q}$ transition~\cite{an1,an2}.

It needs the development from both theory and experiment sides to
understand dynamics of penta-quark states.
New experimental data provide us some clues to extend the existing
baryon spectrum.
As shown in the Particle Group Data (PDG)~\cite{pdg}, the
information of hyperon resonances is very limited.
Recent measurements from CLAS~\cite{guoexp}, LEPS~\cite{LEPS} and
Crystal  Ball (CB)~\cite{CBdata, CBdata2} provide us new information
of $\Sigma^*$ and $\Lambda^*$ resonances.
Correspondingly, analyses of these data~\cite{gao1, Shi, Gao} bring
great changes to understand the properties of $\Sigma^*$ and
$\Lambda^*$ resonances.
All of these new changes strongly support unquenched models.
In order to go beyond the large mixture between three-quark and
five-quark configurations, we used the hidden charm and beauty
components to replace the light $q\bar{q}$ pair in the five-quark
configurations~\cite{Ncc1,Ncc2}.
As a result, several new $N^*$ and $\Lambda^*$ resonances with
hidden charm and beauty are predicted with super-heavy mass and
narrow width.
The super-heavy mass is due to the $c\bar{c}$ and $b\bar{b}$
components and the narrow width stems from the small coupling
between $c\bar{c}$ or $b\bar{b}$ and light quark pairs.
If confirmed, they definitely have penta-quark component dominance.
Very recently, two of such $N^*$ with hidden charm might have been
observed by the LHCb experiment in the decay of $\Lambda
_b$~\cite{LHCb}. More of those states are expected to be observed in
near future. This opens a new window for study hadronic dynamics for
the multi-quark states.

\section{Baryon Spectroscopy with Strangeness}

In Ref. \cite{pdg}, it lists a lot of ground and excited states of
$\Sigma$, $\Xi$ and $\Omega$ baryons.
However, the well established states (marked as four-star) only
include six $\Sigma$ states, two $\Xi$ states and one $\Omega$
state.
Especially, only ground state of $\Omega$ is confirmed as
$\Omega(1670)\frac{3}{2}^+$.
There is no experimental determination on the quantum numbers for
the excited $\Omega$ resonances.
Moreover, the $J^P$ of the lowest excited baryon states in the
quenched quark model is $1/2^-$. However, for these hyperon
resonances, there are no established states with this $J^p$.
On experimental side, the measurement about these hyperon resonances
are all from old experiments of 1970 $-$ 1985 before 2005.
Fortunately,  there are some new observations about $\Sigma^*$ and
$\Lambda^*$ from  CLAS~\cite{guoexp}, LEPS~\cite{LEPS} and  CB group
\cite{CBdata}.
On theoretical side, the unquenched quark model provides totally
different predictions of $1/2^-$ strangeness baryons.
For example, the classical quenched quark models~\cite{qqqCap}
predict the  $1/2^-$ $\Sigma^*$ and $\Xi^*$ to be around 1650 MeV
and 1760 MeV, respectively, while the unquenched quark models expect
them to be around 1400 MeV and 1550 MeV\cite{zounul, 5qRiska,
zhang}, or 1450 MeV and 1620 MeV \cite{oh, khemchandani, ramosprl},
respectively.
It is necessary to check the predictions of hyperon resonances in
these unquenched quark models with new data.
On the other hand, each baryon might be a mixture of the three-quark
and five-quark components. However, the mechanism of transition
between them is not well established.
Here based on the Nambu-Jona-Lasinio (NJL) model and
instanton-induced  interaction, the mass of the lowest $\Omega^*$  is
predicted by including not only five-quark and three-quark
Hamiltonian, but also the potential of transition between
them~\cite{an1, an2}.

\subsection{New analyses of CB data}

In Ref.\cite{CBdata}, differential cross sections and hyperon
polarizations for $\bar{K}^0n$, $\pi^0\Lambda$, and $\pi^0\Sigma^0$
productions in $K^-p$ interactions at eight $K^-$ momenta between
514 and 750 MeV/c were measured.
It provides us a nice place to explore the pure isospin $I=1$ and 0
channels, respectively.
The new combined fit of these new data with old data~\cite{Morris}
on $K^-n \to \pi^- \Lambda$ for the pure I=1  is performed, and for
the pure I=0 channel\cite{Gao}, the new data of $K^- p \to
\pi^0\Sigma^0$  are also analyzed\cite{Shi}.
The fit results of both two channels provide a new spectrum of
$\Sigma^*$ and $\Lambda^*$, which are consistent with predictions of
unquenched models.

In the $\Sigma^*$ sector, new analyses of differential cross
sections and $\Lambda$ polarizations for reactions $K^-n \to
\pi^-\Lambda$ and $K^-p \to \pi^0\Lambda$ are performed by the
effective Lagrangian method, and the experimental data are from the
new high-statistic CB experiment \cite{CBdata} and the early report
of Ref.\cite{Morris}, with the c.m. energy $1550 - 1676$ MeV.
In the analyses, the t-channel $K^*$ exchange and the u-channel
proton exchange amplitudes are fundamental backgrounds.
The well-established four-star $\Sigma(1189)\frac{1}{2}^+$,
$\Sigma^*(1385)\frac{3}{2}^+$, $\Sigma^*(1670)\frac{3}{2}^-$, and
$\Sigma^*(1775) \frac{5}{2}^-$ contributions are always included in
analyses.
If only including above u,t-channel backgrounds and s-channel
resonances, the $\chi^2$ of the best fit arrives 1680 for total 348
data points.
Then each additional resonance of $J^p=$ $\frac{1}{2}^-$,
$\frac{1}{2}^+$, $\frac{3}{2}^-$ and $\frac{3}{2}^+$ reduces the
$\chi^2$ to 899, 572, 943, and 1392, respectively.
Obviously, the data favor a $\frac{1}{2}^+$ $\Sigma^*$ resonance with the mass of 1635 MeV.
It is worthy to mention that polarizations data play the most
important role, which discriminates the
$\Sigma^*(1620)\frac{1}{2}^-$ from $\Sigma^*(1635)\frac{1}{2}^+$.
This analysis shows that $\Sigma^*(1660)\frac{1}{2}^+$ is definitely
needed, while $\Sigma^*(1620) \frac{1}{2}^-$ is not needed at all.
With the further investigation, the $\Sigma^*\frac{1}{2}^-$ with
much  lower mass, as suggested by the penta-quark
model~\cite{5qRiska}, cannot be excluded.
Therefore, there is no evidence for the $\Sigma^*$ with
$\frac{1}{2}^-$ suggested by the quenched quark model around 1600
MeV.
In addition, other $\Sigma$ resonances may exist,
$\Sigma^*(1610)\frac{1}{2}^+$, $\Sigma^*(1542)\frac{3}{2}^-$ and
$\Sigma^*(1840)\frac{3}{2}^+$, where $\Sigma^*(1542)\frac{3}{2}^-$
is consistent with the structure of $\Sigma(1560)$ or $\Sigma(1580)$
resonance in PDG~\cite{pdg}.

In the $\Lambda^*$ sector, the reaction $K^- p \to \pi^0\Sigma^0$,
as a pure I=0 process, can be used to identify the structures of
$\Lambda$ resonances.
As we known, the polarization data is crucial important for analyses.
However, with different data selection cuts and reconstructions, two
groups in the same collaboration, i.e.,  the UCLA group
\cite{CBdata} and the VA group \cite{CBdata2}, gave the very
different polarization data.
By fitting both two sets of the data, it is interesting to find that
dropping four-star $\Lambda(1690)\frac{3}{2}^-$ only increases
$\chi^2 / N$ 0.004, while dropping any other resonance will increase
$\chi^2 / N$ more than 0.5.
In other words, the four-star $\Lambda(1690)\frac{3}{2}^-$ is not
needed here, while at the same energy range $\sim$ 1680 MeV, there
is strong evidence for the existence of a new
$\Lambda^*(\frac{3}{2}^+)$ resonance.
As a result, the contribution of the new
$\Lambda^*(1680)\frac{3}{2}^+$ replaces the contribution from the
four-star $\Lambda(1690)\frac{3}{2}^-$, which has important
implications for hyperon spectroscopy and its underlying dynamics.
The lowest $\Lambda(\frac{3}{2}^+)$ is predicted around 1900 MeV in
the $qqq$ constituent quark model \cite{qqqCap}, which is consistent
with $\Lambda^*(1890)(\frac{3}{2}^+)$ in the PDG.
However, the penta-quark dynamics predicts it to be below 1700
MeV~\cite{5qRiska}, which is corresponding to this new
$\Lambda^*(1680)\frac{3}{2}^+$ here.

By these two new investigations and previous researches about the
$\Sigma^*$\cite{sigma121380} and a new narrow
$\Lambda(\frac{3}{2}^-)$~\cite{LiuXie},  new spectra of $\Sigma^*$
and $\Lambda^*$ are well consistent with those expected in
unquenched quark models.
New $\Sigma(1380)\frac{1}{2}^-$ and $\Sigma(1635)\frac{1}{2}^+$ are
corresponding to the predictions of unquenched quark models,
$\Sigma(\frac{1}{2}^-)$ around 1360 - 1420 MeV and
$\Sigma(\frac{1}{2}^+)$ around 1630 \& 1656 MeV
\cite{MartinezTorres:2008is}, respectively.
On the other hand, there is no evidence for
$\Sigma^*(\frac{1}{2}^-)$ around 1650 MeV as suggested in the
quenched model. For $\Lambda^*$ resonances, the new
$\Lambda^*(1680)\frac{3}{2}^+$ is consistent with the prediction of
Ref.\cite{5qRiska}, and the new
$\Lambda^*(1670)\frac{3}{2}^-$\cite{LiuXie} with narrow width
instead of the broad $\Lambda(1690)\frac{3}{2}^-$ obviously cannot
be explained by quenched quark models.
However, together with the new $\Sigma^*(1542)\frac{3}{2}^-$,
$\Lambda^*(1520)\frac{3}{2}^-$, $N^*(1520)\frac{3}{2}^-$ and either
$\Xi(1620)$ or $\Xi(1690)$, there is a nice $\frac{3}{2}^-$ baryon
nonet with large penta-quark configuration.

It needs a completed low-lying hyperon spectrum to establish the
multi-quark picture for hadronic excited states, especially the
$\frac{1}{2}^-$ and $\frac{3}{2}^-$ $\Sigma^*$, $\Xi^*$ and
$\Omega^*$.
%
%

\subsection{Lowest $\Omega^*$ within $sss \leftrightarrow sssq\bar{q}$}

Five quark components play an important role in the excitation of baryons.
Then the excited baryon may have both $qqq$ and $qqqq\bar{q}$
configurations, which involves the transition between two
components.
The key point of transition is a correct $q\bar{q}$ creation mechanism.
However, in various models, such as $^3P_0$ model\cite{3p0},
string-breaking models\cite{string1} and others\cite{other}, the
$q\bar{q}$ pair creation operator only provides the $^3P_0$ state of
$q\bar{q}$.
However, for low-lying five-quark configurations with the negative
parity, five quarks are all supposed to be relative S-wave.
As a result, these $q\bar{q}$ pair creation operators cannot
contribute to the transition between $qqq$ and $qqqq\bar{q}$.

In recent Refs.\cite{an1,an2}, the instanton-induced interaction and
NJL model are applied for new $q\bar{q}$ pair creation mechanisms,
which create $q\bar{q}$ pairs with quantum numbers $^3P_0$ and
$^1S_0$.
By applying these three-quark and five-quark transitions in the
Hamiltonian matrix, the lowest excitation of $\Omega$ is predicted
to be around 1780 MeV with $\frac{3}{2}^-$.
This result is very different from the prediction of the quenched
model where the lowest $\Omega^*$ is around 2020 MeV with relative
angular momentum L=1 \cite{qqqChao}.
On the other hand, if only with the five-quark components, the
lowest  $\Omega^*$ was predicted around 1820 MeV~\cite{yuan}, and
Ref.\cite{wang1} also predicts the lowest $\Omega^*$ as $\bar{K}\Xi$
bound state around 1805 MeV.
It shows that predicted mass of the lowest $\Omega^*$ in
multi-quarks models is much lighter than that in the three-quark
model.
It is very important to measure where is the lowest $\Omega^*$
experimentally.
Recently, the Beijing Spectrometer II (BESII) collaboration at
Beijing  Electron Positron Collider (BEPC) has already observed the
$\Psi(2S) \to \Omega \bar{\Omega}$ which branch ratio is
$(5\pm2)\times 10^{-5}$ \cite{bes}.
Now with the upgraded BEPC, billions of $\psi(2S)$ events will be
collected by BESIII Collaboration \cite{Asner:2008nq}, which is two
orders of magnitude higher than what BESII experiment got.
The mass  upper limitation of $\Omega^*$ in $\Psi(2S) \to \Omega^* \bar{\Omega}$  is 2030 MeV.
So it is a nice place to examine the existence of the $\Omega^*$
resonance predicted by the multi-quark models.
Once the lowest $\Omega^*$ is fixed, there will be a clear picture
for the internal structure of $\Omega^*$ states.

\section{From Strangeness to Charm and Beauty}

As discussed in the introduction, a lot of well established $N^*$
and $\Lambda^*$ resonances were proposed to have large five-quark
configurations, such as $N^*(1535)$ and $\Lambda^*(1405)$.
However, they are hard to distinguish from classical quark model
states due to tunable ingredients and possible large mixing of
various configurations in these models.
In the PDG-2010~\cite{pdg2010}, it still claimed:
``The clean $\Lambda_c(2595)$ spectrum has in fact been taken to
settle the decades-long discussion about the nature of the
$\Lambda(1405)$ -- true 3-quark state or mere $\bar{K}N$ threshold
effect -- unambiguously in favor of the first interpretation."
Obviously, this claim is not justified, and it disappears in the
later versions of PDG~\cite{pdg}.
Actually, Refs.\cite{lambda25951,lambda25952} propose the
$\Lambda_c(2595) \frac{1}{2}^-$ to be $DN$ molecule.

Now the question is that how to distinguish three-quark and
five-quark components in baryons. It is too difficult for us to
answer.
Then making the question simpler: where is the five-quark dominant
state?
Here we only consider such $q_1q_2q_3q\bar{q}$ state  where $q$ and
$\bar{q}$ have the same flavor.
Thus the exotic penta-quark states are not discussed here, such as
$\theta^+$ with $|udud\bar{s}>$, since up to now there is no
convincing evidence for these states.
A possible alternative solution of the five-quark dominant state is
to extend lighter quark-antiquark pair to the heavy quark-antiquark
pair.
If baryons have the heavy $Q\bar{Q}$ ($Q=c, b$) components, their
masses will be definitely much larger than ordinary baryons composed
of three light quarks, while with much narrower width.
Therefore, such super-heavy baryons are definitely beyond the naive
$qqq$ consistent quark model.
For example, if the $N^*(1535)$ is the $\bar{K}\Sigma$ quasi-bound
state or $[ud][us]\bar{s}$ diquark-diquark-antiquark state with
hidden strangeness, naturally, by replacing $s\bar{s}$ as $c\bar{c}$
or $b\bar{b}$, some super-heavy $N^*$ states with hidden charm or
beauty may exist.
The possible states would be strongly coupled with $\bar{D}\Sigma_c$
and $B\Sigma_b$ channels, respectively.
In the first subsection, the predictions of such $N^*$ and
$\Lambda^*$ resonances will be introduced.
In the second subsection, we will introduce the study for searching
baryons with hidden charm and beauty in experiments.
%

\subsection{Predictions of $N^*$ and $\Lambda^*$ resonances with hidden charm and beauty}

By following the Valencia approach~\cite{oset1}, the model is
extended from three flavors to four.
Refs.\cite{Ncc1,Ncc2} consider two sets of coupled-channels,
pseudoscalar-baryon ($PB$) channels including $\bar{D} \Sigma_c$,
$\bar{D}\Lambda_c$, and $\eta_c N$, and vector-baryon ($VB$)
channels including  $\bar{D}^* \Sigma_c$, $\bar{D}^*\Lambda_c$, and
$J/\psi N$.
With the interaction by exchanging vector meson, the T matrix of $PB
\to PB$ and $VB \to VB$ can be obtained by solving the coupled
channels Bethe-Salpeter (BS) equation in the Valencia approach of
Ref.~\cite{oset1}.
Then we look for poles of T matrix in the complex plane of
$\sqrt{s}$.
If pole appears in the first Riemann sheet below threshold, it is
considered as bound states whereas it located in the second Riemann
sheet and above the threshold of some channels is identified as
resonance.
This meson-baryon model dynamically generates six narrow states from
$PB$ and $VB$ channel:  two $N^*$ resonances and four $\Lambda^*$
resonances .
As shown in Tab. \ref{tab:heavy}, all of these resonances are with mass around 4.3 GeV and width smaller than 100 MeV.
In the Valencia approach,  a static limit is assumed, which leads to neglect spin and momentum dependent terms of interaction potential.
Therefore, only S-wave is considered here.
The predicted resonances from $PB$ channels have $J^P =
\frac{1}{2}^-$, while from $VB$ channels there are degenerated  $J^P
= \frac{1}{2}^-$ and $\frac{3}{2}^-$ states.
Obviously, these super-heavy $N^*$ and $\Lambda^*$ resonances have
nearly pure $qqqc\bar{c}$ ($q = u$, $d$ or $s$) components because
they are all the qusi-bound states of anti-charmed meson-charmed
baryon with negligible couplings to channels without charm.

\begin{table}[ht]
 \caption{Pole position ($Z_R$), mass M and total width $\Gamma$ (including the contribution from the light meson and baryon channel). The units are in MeV.}
 \begin{center}
\begin{tabular}{cccccccc}\hline\hline
&\multicolumn{3}{c}{$PB$ channel}  && \multicolumn{3}{c}{$VB$ channel}   \\
\hline
$(I, S)$         & $Z_R$                &M             &  $\Gamma$       &  &$Z_R$            & M          & $\Gamma$       \\
\hline
$(1/2, 0)$     & $4265-11.6i$     & $4261$   & $56.9$             &  &$4415-9.5i$   & $4412$ & $47.3$\\
\hline
$(0, -1)$       & $4210-2.9i$     & $4209$     & $32.4$             &  &$4547-2.8i$   & $4368$ & $28.0$\\
                     & $4398-8.0i$     & $4394$     & $43.3$             &  &$4368-6.4i$   & $4544$ & $36.6$\\
\hline
\end{tabular}
\end{center}\label{tab:heavy}
\end{table}

To investigate the possible influence of the assumption of potential
in the Valencia approach, Ref. \cite{Ncc3} uses EBAC approach to
re-check the prediction of baryons with hidden charm.
In this approach, the T matrix is solved from the three dimensional
scattering equation which is a reasonable assumption of BS equation.
It is benefit to avoid any assumption of the potential.
In this calculation, the $N^*$ and $\Lambda^*$ resonances are also
dynamically generated although the mass and width of them are
slightly different.

On the other hand, by replacing $c\bar{c}$ with $b\bar{b}$ and using
the same meson-baryon model with the Valencia approach, two $N^*$
resonances and four $\Lambda^*$ resonances with hidden beauty are
dynamically generated.
Because of the super heavy $b\bar{b}$ pairs involved in these
states,  masses of them are all around 11 GeV while widths are only
a few MeV.
In order to study the uncertainties from the assumption of the
Valencia  approach, especially from the momentum dependent terms, we
also used the conventional Schroedinger Equation approach to confirm
the $N^*$ with hidden beauty from $B\Sigma_b$ channel \cite{Nbb}.
The consistent result gives some justifications of the simple Valencia approach.
Before the new observations from the LHCb collaboration~\cite{LHCb},
there were a lot of predictions about these super-heavy states with
hidden charm in other meson-baryon models, with masses above $J/\psi
p$ threshold~\cite{uchino, wang, yang} in consistent with
ours~\cite{Ncc1,Ncc2} although there were some earlier predictions
with masses below $J/\psi p$ threshold~\cite{Hofmann,gobbi}.
It shows that a series of super-heavy $N^*$ and $\Lambda^*$
resonances possibly exist around 4.3 GeV and 11 GeV in various
meson-baryon scattering models.

Unlike above meson-baryon scattering models, penta-quark state
$|qqqq\bar{q}>$ can be also consisted of the colored quark cluster
$[qq][qq]$ and $\bar{q}$.
Ref. \cite{yuancc} uses three kinds of schematic interactions, the
chromomagnetic interaction, the flavor-spin dependent interaction
and the instanton-induced interaction, to study low-lying energy
spectra of penta-quark system with $uudc\bar{c}$ and $udsc\bar{c}$.
The lowest penta-quark state has an S-wave orbital angular momentum
and $J^P=1/2^-$, and they are predicted with the mass around 4.1
GeV.
The interesting thing is that here the predicted lowest mass of
$udsc\bar{c}$ state is heavier than the $uudc\bar{c}$ state, because
the strange quark is heavier.
However, as shown in Tab.~\ref{tab:heavy}, the lowest $\Lambda^*$
resonance is lighter than the $N^*$ resonance in the meson-baryon
scattering model, because the threshold of $\bar{D}_s \Lambda^+_c$
is below that of $\bar{D} \Sigma_c$.
The different mass order between $N^*$ and $\Lambda^*$ resonances
can be used to distinguish these two models in the future.

\subsection{Experiment evidence and further exploration}

Just after this conference, two states $P^+_c(4380)$ and
$P^+_c(4450)$ were claimed to be observed in the invariant mass
spectrum of $J/\psi p$ in decay reaction $\Lambda_b \to J/\psi K^-
p$ by the LHCb Collaboration \cite{LHCb}.
Their masses are found to be $4380\pm 8\pm29$ MeV and
$4449.8\pm1.7\pm2.5$ MeV, with corresponding widths of
$205\pm18\pm86$ MeV and $39\pm5\pm19$ MeV, respectively.
The preferred $J^P$ assignments are of opposite parity, with one
state having spin $\frac{3}{2}$ and the other $\frac{5}{2}$.
This new observation attracts a lot of theoretical interests.
There are three different views of these two new states: a)
anticharmed meson-charmed baryon molecular states~\cite{chenpa,
chenpb, roca, he}, b) penta-quark states consisted of colored quark
cluster based on diquark models~\cite{lebed,Maiani:2015vwa}, and c)
a kinematical effect for one peak~\cite{guo, liu}.

Obviously, the first and the second views which both regard these
two states as multi-quark states are consistent with previous
Refs.\cite{Ncc1,Ncc2,Ncc3,uchino, wang, yang} and Ref.\cite{yuancc},
respectively.
In order to confirm whether they are genuine physical states or just
some kinematical effects, as well as to find other such heavy states
with different quantum numbers, new experiments for these
super-heavy states is essential.
In Refs.\cite{Ncc1,Ncc2,Ncc3, hejun,Wang:2015jsa}, the reaction
$\gamma p \to J/\psi p$ has already been proposed to look for the
$N^*$ within hidden charm, possibly at the CEBAF-12 GeV-upgrade in
Jefferson Lab.
Furthermore, other possible channels $\eta_c p$ and $\Upsilon p$ are
also suggested for searching the new $N^*$ resonance within hidden
charm and beauty, respectively.
For the PANDA/FAIR, with a $\bar{p}$ beam of 15 GeV one can get the
total energy of $\bar{p}p$ collisions arrive 5470 MeV, which allows
one to observe $N^*$ resonances in $pX$ production up to a mass
$M_X\sim 4538$ MeV or a $\Lambda^*$ resonances in $Y\Lambda$
production up to a mass $M_Y\sim  4355$ MeV.
It will be a nice place to examine the existence of super-heavy
$N^*$ and $\Lambda^*$ resonances. These super-heavy $N^*$ and
$\Lambda^*$ could also be looked for from $\pi p$ and $Kp$
experiments~\cite{Wang:2015qia,Wang:2015xwa}, possibly at JPARC.
And for the $N^*$ with hidden beauty, the available center-of-mass
energies of $pp$ and $ep$ collisions are larger than 13 and 14 GeV,
respectively, and cross sections of $pp \to pp\Upsilon$ and $e^- p
\to e^- p \Upsilon$ should be larger than 0.1 nb.
It is expected new facilities in future, such as proposed
electron-ion collider~\cite{zounul}.

\section{Conclusions}

With more and more hadronic states being discovered in experiments,
the quenched quark models become too simple to explain properties of
various hadrons, while the unquenched quark models are needed.
Comparing to the orbital angular momentum excitation mechanism in
quenched quark models, the $q\bar{q}$ dragged out from gluon field
is another important excitation mechanism for hadrons.
It is necessary to go beyond the classical quenched quark model: the
number of quarks in a hadron is not a constant.
New experimental observations of the CB group play a key role for
understanding hyperon spectrum, and new analyses of these data
strongly support unquenched quark model.
Furthermore, based on the transition between $sss$ and $sssq\bar{q}$
with the NJL model and instanton-induced interaction, the new lowest
$\Omega^*$ is predicted around 1780 MeV.
It is expected to be observed in the $\psi(2S) \to \bar{\Omega}\Omega^*$
reaction at BESIII.
In order to avoid the mixture between five-quark and three-quark
components in baryons, we extend several models of baryons to the
cases with hidden charm and beauty.
Then a series of super-heavy $N^*$ and $\Lambda^*$ resonances with
hidden charm and  beauty are predicted to be around 4.3 GeV and 11
GeV from various unquenched quark models.
Fortunately, two of such $N^*$ with hidden charm might have been
observed by the LHCb experiment.
Confirmation from other experiments and some further detailed
investigations of these new $N^*$ resonances from both experimental
and theoretical sides are essential to build up penta-quark
dynamics.
Searching for their partners of various quantum numbers are also
crucial for understand the multi-quark dynamics.

$\\$
\textbf{Acknowledgment}
$\\$

We thank C.S. An, P.Z. Gao, J. He, F. Huang, T.S.H. Lee, R. Molina,
E. Oset, J. Shi, W.L. Wang, K. W. Wei, H.S. Xu, S.G. Yuan, L. Zhao,
Z.Y. Zhang for collaboration works reviewed here. This work is
supported by the National Natural Science Foundation of China under
Grant 11261130311 (CRC110 by DFG and NSFC), the Chinese Academy of
Sciences under Project No.KJCX2-EW-N01.

\end{document}